# Interference of diffraction and transition radiation and its application as a beam divergence diagnostic

R. B. Fiorito and A.G. Shkvarunets
IREAP, University of Maryland, College Park, MD

T. Watanabe and V. Yakimenko
ATF, Brookhaven National Laboratory, Upton, NY

D. Snyder
Dept. of Physics, Naval Postgraduate School, Monterey, CA

Abstract

We have observed the interference of optical diffraction radiation (ODR) and optical transition radiation (OTR) produced by the interaction of a relativistic electron beam with a micromesh foil and a mirror. The production of forward directed ODR from electrons passing through the holes and wires of the mesh and their separate interactions with backward OTR from the mirror are analyzed with the help of a simulation code. By careful choice of the micromesh properties, mesh-mirror spacing, observation wavelength and filter band pass, the interference of the ODR produced from the unperturbed electrons passing through the open spaces of the mesh and OTR from the mirror are observable above a broad incoherent background from interaction of the heavily scattered electrons passing through the mesh wires. These interferences (ODTRI) are sensitive to the beam divergence and can be used to directly diagnose this parameter. We compare experimental divergence values obtained using ODTRI, conventional OTRI, for the case when front foil scattering is negligible, and computed values obtained from transport code calculations and multiple screen beam size measurements. We obtain good agreement in all cases.

**Introduction**

The term 'diffraction radiation' is commonly used to describe the radiation produced when a charged particle moving at a constant velocity passes near, but does not intercept, a material whose dielectric constant differs from the medium in which the particle is traveling [1]. This radiation is caused by a rapid change in the induced polarization of the impacted medium caused by the transiting particle. It is the polarization current that radiates. The radiation can be observed in the far field (Fraunhofer zone) or the near field (wave or Fresnel zone). The far field spectral-angular properties of DR are similar to those of transition radiation (TR), which is produced by a charge particle *passing through* a solid boundary between two media with different dielectric constants, but has some distinguishing features [2]. Like TR, the spectral-angular distribution of DR is altered by the angular distribution of the beam particles and thus it can be used to diagnose the beam's divergence and mean trajectory angle. However, unlike TR, the spectral-angular distribution of DR is also a function of the beam size and its intensity is a function of the proximity of the charged particle to the impacted medium.

The relevant parameter which governs the intensity of DR produced is the so-called radiation impact parameter, $a = \gamma\lambda/2\pi$; here $\gamma$ is the Lorentz factor of the moving charge and $\lambda$ is the observed wavelength. The parameter $a$ is a measure of the degree of fall-off of the radial component of electric field of the moving particle [3]. Significant DR is produced when $a \leq l$, the distance of closest approach of the particle to the impacted medium, e.g. the edges of circular aperture or slit through which the beam traveling in a vacuum passes. The radiation impact parameter is also relevant to the production of transition radiation, since when $a \geq r$, the size of the radiating medium, diffraction effects from the edges of the radiator are significant [4,5]. These effects include cutoffs in the spectral density at low frequency for a finite size solid radiator and at high frequency for an aperture, as well as modulations (fringes) in the angular distribution of the radiation.

TR from a finite size screen and DR from an aperture are closely related complementary effects. In fact, Babinet's principle applies to the radiation fields; i.e. TR from a finite size radiator is equal to the difference of TR from an infinite plane and DR from a complementary aperture [6-8]. In this sense there is no formal distinction between the two radiation phenomena and *we will refer to both TR from a finite size screen and DR from an aperture as diffraction radiation, when the relevant size of the radiator or aperture is less than or of the order of the radiation impact parameter.* Our interest in this paper is the investigation of incoherent ($\lambda <<$ beam dimensions), far field, optical ($\lambda$ = 400-700 nm) diffraction radiation (ODR) from beams with moderate energies (i.e. 10- 100 MeV), where the radiation impact parameter $a$ is in the range 10-100 μm.

Theoretical investigations have shown that the far field angular distribution (AD) of ODR can be used as a non-interceptive beam size and divergence diagnostic for relativistic beams [6,9,10]. Experiments have verified that the far field AD from a single screen can be used to measure the beam size for a low divergence beam [11]. '*Near field imaging*', a term which is somewhat loosely used to describe imaging the spatial distribution of DR at the source to elicit information about beam position and size has also been investigated theoretically [12,13]. Recently an experimental study of 'near field imaging' of ODR from a single metal edge has been reported [14].

In an earlier study [7] we showed computationally how optical diffraction-transition radiation interferometry (ODTRI) could be used to measure the divergence of moderate energy (10-100 MeV) electron beams. This technique uses a device similar to a conventional two foil OTR interferometer [15]. However, in an ODTR interferometer the first foil is replaced by a micromesh [16] whose cell dimensions are comparable to the radiation impact parameter for visible wavelengths but much smaller than the beam radius (100's to 1000's of microns).

A general schematic of a reflection ODTR interferometer is shown in Figure 1. The diagram shows the production of ODR from unscattered (u) electrons passing through the holes of the micromesh and ODR from scattered (s) electrons passing through the wires of the mesh. These forward directed ODR components reflect and interfere with backward OTR generated by the beam impinging on the mirror itself.

While both ODR components from the mesh are due to a changing induced polarization current on the metal in wires we can consider them to be independent effects. The total ODR intensity from unperturbed electrons passing through the holes is comparable to that produced by electrons passing through the wires when the transparency of the mesh is about 50% [7]. However, by properly choosing the atomic number and thickness of the mesh material one can take advantage of electron scattering in the wires to wash out the interference of ODR from the scattered component and backward OTR from the mirror produced by this component. The scattered contribution to the observed radiation pattern then forms a smooth background and by proper choice of the optical band pass, the fringes from the ODR from *unperturbed* electrons passing through the holes can be made visible above this background. The visibility of these fringes is sensitive to the unperturbed beam divergence.

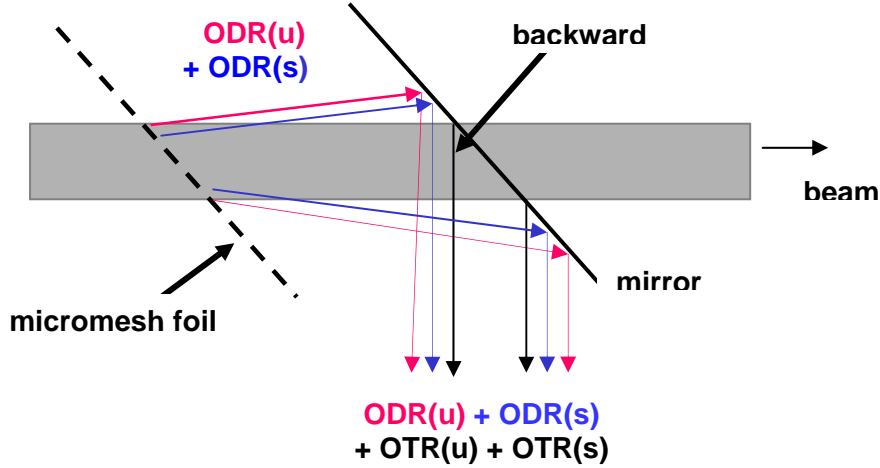

Figure 1. Schematic of the ODTR Interferometer showing various radiation components.

In [17] we presented preliminary results of an rms beam divergence measurement made using ODTRI. In this paper we present detailed results, analysis and comparisons of divergences obtained using three different techniques: ODTRI, OTRI and multiple screens-transport code calculations. We report measurements of both vertical and horizontal components of the divergence on two different electron beam accelerators with beam energies 50 and 100 MeV, respectively. We also present a more detailed explanation of the model employed in the simulation code we have developed to calculate ODTRI than previously given in ref. [17]. We also provide a detailed explanation of how we use the simulation code results to fit the data. The excellent agreement between all these measurements and calculations firmly establishes ODTRI as a viable new technique for the measurement of beam divergence for moderate energy relativistic electron beams. In addition, ODTRI has a distinct advantage over conventional OTR interferometry, which is subject to the limitation that only divergences comparable to or exceeding the rms scattering angle in the primary foil can be measured; no such limitation is present with ODTRI.

**Background**

OTR Interferometry

The performance of a conventional OTR two-foil interferometer can be evaluated from the expression for the far field spectral-angular distribution of backward reflected radiation observed in the detection plane. This plane is perpendicular to the direction of specular reflection (for backward reflected radiation) or to the direction of the average beam velocity (for forward radiation). While in reality the radiation expands as a spherical wave in the far field, measurement of the radiation in the detection plane, which is tangent to the spherical wave front, is a good approximation for small angles of observation measured from the tangent point, i.e. z = 0.

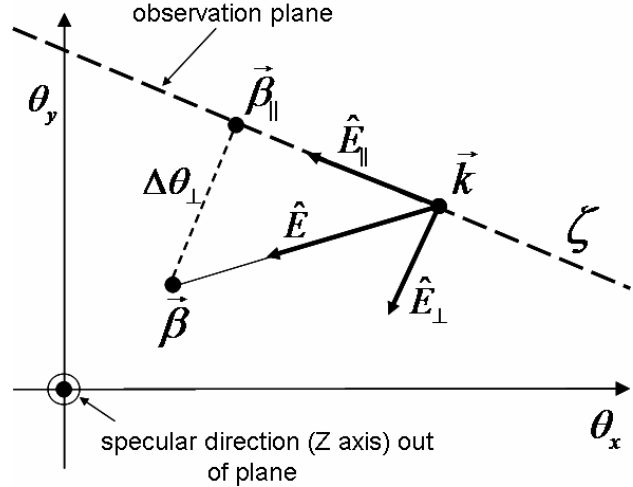

Figure 2. Schematic of the detection plane.

We introduce spherical angles $\theta_x$ and $\theta_y$ to describe the radiation measured in the detection plane which is shown in Figure 2. In this plane the positions of vectors are represented as points and planes intersecting the detection plane are represented as lines joining two vectors. Shown in the Figure are the vectors **k**, the radiation wave vector, $\boldsymbol{\beta} = \mathbf{V}/c$, where |**V**| is the beam velocity and the direction of **V** is directed along z, i.e. the direction of specular reflection, $\boldsymbol{\beta}_{\|}$ is the component of β parallel to the observation plane $(\mathbf{k}, \boldsymbol{\beta}_{\|})$, $\hat{\mathbf{E}}$ is the electric field of the radiation, $\hat{\mathbf{E}}_{\|,\perp}$, are the ∥ and ⊥ components

of $\hat{\mathbf{E}}$ with respect to the observation plane and $\zeta$ is the observation or scan angle in the observation plane measured from the direction of $\boldsymbol{\beta}_\parallel$. Note that the observation plane can be oriented arbitrarily in the detector plane and, in general, it does not pass through the direction $\boldsymbol{\beta}$ nor through the z axis. Note also that $\boldsymbol{\beta}$ is not generally collinear with the z axis.

The far field spectral-angular density for *interference OTR* measured in the observation plane is given by:

$$\frac{d^2 I_{\parallel,\perp}^{int}(\omega,\zeta)}{d\omega d\Omega} = \left|r_{\parallel,\perp}\right|^2 I_{\parallel,\perp}(\zeta)\left|1-e^{-i\Psi}\right|^2 \qquad (1)$$

where $\omega$ is the frequency, $\Omega$ is the solid angle subtended by the source at the detector plane, $\left|r_{\parallel,\perp}\right|^2$ are the $\parallel,\perp$ Fresnel reflection coefficients of the foil, which are both approximately unity for a highly conductive metallic surface, $I_{\parallel,\perp}(\zeta)$ are the single foil OTR intensities polarized parallel and perpendicular to the plane of observation and $\Psi$ is the phase difference between the radiations generated at the two foils [15].

The intensities $I_{\perp,\parallel}(\zeta) \propto \left|\hat{\mathbf{E}}_{\perp,\parallel}(\zeta)\right|^2$ in the observation plane are symmetric around $\boldsymbol{\beta}_\parallel$ and for angles close to $1/\gamma$ are given by:

$$I_\parallel(\zeta) = \beta^2 \frac{e^2}{\pi^2 c} \frac{\zeta^2}{(\gamma^{-2}+\beta_\perp^2+\beta^2\zeta^2)^2}, \qquad (2)$$

$$I_\perp(\zeta) = \beta_\perp^2 \frac{e^2}{\pi^2 c} \frac{1}{(\gamma^{-2}+\beta_\perp^2+\beta^2\zeta^2)^2} \qquad (3)$$

where $\beta_\perp$ is the amplitude of the perpendicular component of $\boldsymbol{\beta}$ and e is the charge of the electron. Note that in the special case when $\boldsymbol{\beta}$ is in the observation plane, $\beta_\perp = 0$ and $I_\perp = 0$ for OTR. If, in addition, the direction of $\boldsymbol{\beta}$ is collinear with the z axis and $\beta \approx 1$, Eqs. (1,2) assume the forms most often seen in texts and papers on TR:

$$\frac{d^2I^{int}(\theta)}{d\omega d\Omega} = 4\,I(\theta)\sin^2(d/2L_V) \tag{4}$$

and

$$I(\theta) = I_\parallel(\theta) = \frac{e^2}{\pi^2 c}\frac{\theta^2}{(\gamma^{-2}+\theta^2)^2}, \tag{5}$$

where the *sine* term of Eq. (4) represents the interference of two sources separated by distance *d* and $L_V = (\lambda/\pi)(\gamma^{-2}+\theta^2)^{-1}$ is the coherence or 'formation length', defined as the distance over which the field of the electron and the co-moving radiation photon differ in phase by 1 radian.

For all inter foil distances the radiation from two foils will interfere. However, the number of interferences per angular interval increases as the interfoil spacing and angle of observation increase. The visibility of these interferences is a function of the divergence and energy spread of the beam which typically are fractions of 1/γ for high quality beams. However, we have shown [21] that if the energy spread is smaller than the normalized divergence of the beam (i.e. ΔE/E << γσ), which is the case for our experimental conditions for all angles, the divergence effect dominates and hence the fringe visibility becomes a diagnostic for this quantity. The sensitivity of the interferometer to a given range of divergence can be optimized by adjustment of the interfoil spacing and the band pass of the measurement.

**Description of Simulation Code used to Calculate ODR and ODTRI**

In a conventional OTR interferometer forward OTR from a solid foil reflects and interferes with backward OTR from the mirror, where both the forward TR from the first foil and the backward TR from the mirror each has the form given above in Eqs. (2,3,5). However, when the first foil is a mesh, the radiation is ODR from two distinguishable sources: 1) the beam electrons passing through the holes of the mesh and 2) the beam electrons passing through the solid wires separating the holes. Each of these ODR components interferes with two corresponding OTR components generated from the unscattered and scattered beam electrons emerging from the mesh interacting with the mirror. Thus the total intensity observed is composed of four contributions, two ODR components from the mesh and two corresponding OTR components from the mirror.

Analytic expressions for $I_{\parallel,\perp}^{ODR}$ from the mesh similar to $I_{\parallel,\perp}^{OTR}$ given above in Eqs. (2,3) are not available. Hence we have developed a simulation code to calculate them. In addition our code computes the angular convolution of these components with a two dimension distribution of beam trajectory angles represented by one or more 2D Gaussian distributions each with a width representing $\sigma_{x,y}$, the rms *x* and *y* divergences of the corresponding beam component. For OTRI such a convolution can be directly applied to the analytic forms for the OTR intensities components given above by Eqs. (2,3). For ODTRI the convolutions are incorporated into the ODR simulation code. The latter simulation code results should agree in the OTRI calculations in the limit of zero mesh cell size (i.e. continuous foil limit). We have use this limit as well as other checks [see ref. 7] to establish the validity of the simulation code.

*Calculation of DR and TR from the two foils of the interferometer*

Our simulation code calculates the angular distribution of the intensity of ODR produced by an electron beam passing through two parallel foils which are separated by the distance *d* measured along the direction of the beam velocity. In the analysis and experiments described in this paper the foils are tilted by an angle $\nu = 45^O$. In the code we neglect the longitudinal component of the electric field of the electron. This simplifying approximation is valid for high energies (*E* > 50 MeV) even if the foils are tilted with respect to the electron beam velocity. We assume that the mesh perforations are perfectly symmetric rectangular holes with width *h* , which are evenly and symmetrically distributed on the foil with period *p*. The foil structure is represented as sum of translations of the unit cell shown in Figure 3., which shows a portion of the perforated foil projected onto a plane normal to the mean beam velocity. We refer to this plane as the *source plane*. The perforations are shown as white rectangles and a single perforation and its surrounding solid area (unit cell) is shown as two concentric rectangles. Radiation from the first and second foils is calculated assuming that the forward and backward radiations are symmetric about the surface of the tilted foil. If the size of the foil is large and the beam cross section is much larger than the period of the perforations *p*, the beam density profile varies very slowly over the cell period *p* and is considered to be constant over each cell.

For computational purposes, the beam passing through the perforated foil is split into two fractions: one fraction composed of electrons passing the solid part of foil cell (i.e. a scattered component ) shown in dark grey in Figure 3., the other composed of unscattered electrons passing through the holes shown as the lighter grey rectangle. The beam's spatial distribution is modeled as a large number of macro particles ($N \sim$ 1000-2000) which are homogeneously distributed within the cell. The number of unscattered particles $N_U = N \cdot T$, where *T* is the foil transparency and the number of scattered electrons $N_S = N - N_U$.

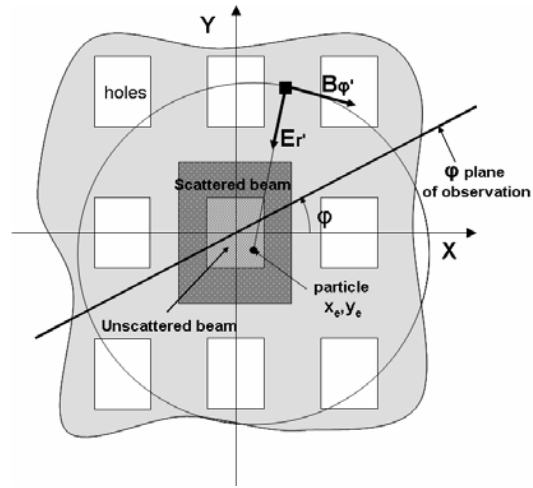

Figure 3. Schematic of the mesh foil projected into a plane perpendicular to the z direction showing unit cell and the region of influence of the field of one electron shown by the circle.

*Calculation of ODR Intensities for a single particle within cell of the mesh*

For simplicity, the formulas presented below describe forward directed radiation from both foils of the interferometer considering the z axis to be directed forward and the detection plane is normal to this axis. There is no loss of generality in this approach because forward and backward specularly reflected radiations are mirror symmetric.

Following the picture introduced in the above paragraph, we introduce Cartesian coordinates $x, y, z$ to describe the mesh perforations and the coordinates $x_e, y_e$ (with radius vector $\vec{r}_e$) to describe the position of the electron in the source plane, *viewed now as normal to the forward direction, and correspondingly $z$ is now directed in the forward direction.*

We introduce various observation planes, which are normal to the source plane. The *horizontal observation plane* is defined to be coplanar to $(x, z)$; the *vertical observation plane* is defined to be coplanar to $(y, z)$; and we use cylindrical coordinates $r, \varphi, z$, where $x = r \cos\varphi$, $y = r \sin\varphi$ to describe the fields in the $\varphi$ *observation plane*, which is a plane perpendicular to $(x,y)$, passing through z and oriented at angle $\varphi$ with respect to the x axis. We also use *local cylindrical coordinates* $r', \varphi', z'$, where $z'$ is parallel to the velocity $\vec{V}$, to describe the fields of the electron in the $\varphi$ *observation plane*. We assume that the electron's trajectory is parallel to the axes $z$, where $z' = z$, $x - x_e = r' \cos\varphi'$ and $y - y_e = r' \sin\varphi'$.

In local cylindrical coordinates $r', \varphi', z'$ the longitudinal Fourier components with respect to time of the electric and magnetic fields of a relativistic electron in free space can be written as:

$$E'_{r'}(r', \varphi', z', \omega) = E'(r', \omega) \exp(i\omega z / V) \qquad (6)$$

$$B'_{\varphi'}(r', \varphi', z', \omega) = B'(r', \omega) \exp(i\omega z' / V) \qquad (7)$$

Note that the electric field has only a radial component, the magnetic field has only an azimuthal component and that both fields are azimuthally symmetric about the $z$ axis. $E'(r', \omega) = e\alpha K_1(\alpha r') / \pi V$ and $B'(r', \omega) = \beta e\alpha K_1(\alpha r') / \pi V$, where $K_1(\alpha r')$ is the MacDonald function of first order, $e$ is the charge of the electron and $\alpha = \omega / V\gamma$. The Fourier components of fields of the electron can be interpreted as waves propagating along with the moving electron whose field is concentrated within a radius $r' \sim 1/\alpha = V\gamma / \omega$.

Now consider an electron which is incident on or emerges from the surface of perfect conductor. Inside the conductor the total field equals zero because the perfect conductor "screens out" all fields. This means that the conductor can be modeled as a region where, in addition to the fields of the electron, there are "primary induced" electric $E_i$ and magnetic $B_i$ fields with amplitudes equal and opposite in sign to the fields of the electron, i.e. $Ei \equiv -Ee$ and $Bi \equiv -Be$ at all points in the conductor including the surface. Additionally we assume that these primary induced fields are " non radiative" inside this region. As a result we can consider the metallic boundary as a surface $S'$ with a known distribution of electric and magnetic source fields.

We assume that the induced surface fields radiate into the vacuum and that the field radiated into free space can be found using the Huygens-Fresnel principle. For example, the components of the electric field parallel and perpendicular to the $\varphi$ plane at the observation point $\vec{R}$ are given by:

$$E_{\parallel,\perp}(\vec{r}_e,\vec{R}) = \frac{k}{2\pi i}\int_{S'} \frac{a_{\parallel,\perp}\cos\nu \cdot \exp(ikR')}{R'}dS' \qquad (8)$$

where $r',\varphi',z'$ are coordinates of the surface element $dS' = r'dr'd\varphi'/\cos\nu$, $k = \omega/c =$ is the modulus of the wave vector $\vec{k} = (k_x,k_y,k_z)$, $\omega$ is the frequency of the radiation, $c$ is the speed of light in vacuum, $a_{\parallel} = E'(r',\omega)\cdot\exp(ikz'/\beta)\cdot\cos(\varphi-\varphi')$, $a_{\perp} = E'(r',\omega)\cdot\exp(ikz'/\beta)\cdot\sin(\varphi-\varphi')$ are the complex amplitudes of the components of the electric field on the tilted surface $S'$, $\vec{R}$ is the radius vector of the observation point and $R'$ is the distance from surface element $dS'$ to the observation point.

Whether the radiation is DR or TR depends only on the size and structure of the area of integration $S'$. The radiation is TR if the area of integration is large (i.e. $\max(r') \geq 10/\alpha$) and the surface is solid, i.e. there are no zeroes of $a_{\parallel,\perp}$ on the area; the radiation is DR if $\max(r') \leq 10/\alpha$ or if there are regions in the area where $a_{\parallel,\perp} = 0$, e.g. holes in the foil where the primary induced fields are zero.

At large distances from the radiator $R' \gg \sqrt{S'}$,

$$E_{\parallel,\perp}(\vec{r}_e,\vec{R}) = \frac{\exp(ikR)}{R} \cdot \frac{k}{2\pi i} \int_{S'} a_{\parallel,\perp} \cdot \cos\nu \cdot \exp(ik \cdot \Delta r) dS' \qquad (9)$$

where $\Delta r = R' - R$. Thus the angular distribution of the radiation is determined by the term

$$\hat{E}_{\parallel,\perp}(\vec{r}_e,\vec{k}) = \int_{S'} a_{\parallel,\perp} \cdot \cos\nu \cdot \exp(ik \cdot \Delta r) dS' \qquad (10)$$

which gives the radiation field produced from the area $S'$ in the direction $\vec{k}$. The spectral energy density at the observation point averaged over the period of oscillation of the field is given by:

$$\frac{d^2W}{d\omega ds} \sim \frac{\hat{E}_{\parallel,\perp}^2(\vec{r}_e,\vec{k})}{4\pi R^2} \qquad (11)$$

where $ds$ is an elementary surface normal to $\vec{k}$ at distance $R$.

In the far field zone (radiation zone) the energy spectral density per unit solid angle $d\Omega$ in the direction $\vec{k}$ ( later in this paper referred to as the intensity of radiation) is:

$$I_{\parallel,\perp}(\vec{r}_e,\vec{k}) = \frac{d^2W}{d\omega d\Omega} = \frac{e^2\gamma^2}{4\pi^2 c} \frac{\hat{E}_{\parallel,\perp}^2(\vec{r}_e,\vec{k})}{\hat{E}_{OTR}^2(\theta_{max})} \qquad (12)$$

where $\hat{E}_{OTR}(\theta_{max}) = \int_{S'_\perp} a_{OTR} \cdot \exp(ik \cdot \Delta r) dS'_\perp$ is calculated at the angle $\sin\theta_{max} = \gamma^{-1}\beta^{-1}$ which corresponds to the peak of OTR intensity at normal incidence, $a_{OTR} = E'(r',\omega) \cdot \exp(ikz'/\beta) \cdot \cos\varphi$ and $S'_\perp$ is a large solid area of integration normal to the particle trajectory.

The angular distribution of DR depends strongly on the size of integration area, the coordinates of the particles, the distribution of the holes in the foil and the angle of inclination of the foil. Note also that, in general, the perpendicular intensity $I_\perp^{DR}$ *is not* *zero even if* **V** *is parallel to the observation plane.* In contrast to DR, the angular distribution of TR is independent of the coordinates, the spatial distribution of the particles and the angle $\varphi$ and $I_\perp^{TR} = 0$, when **V** is parallel to the observation plane.

In our model a small deviation of the trajectory angle of an electron from the z axis corresponds to a small deviation of the tilt angle of the foil from the angle $\nu = 45^0$. We have found that for small angular deviations, i.e. $\Delta \nu \leq 5/\gamma \approx 0.05 \, rad$ the angular pattern of the radiation produced from any particle in the unit cell is practically unaffected by the deviation angle (i.e. the intensity changes less than few percent in the worst case). We conclude that the pattern of radiation of an electron deflected from the z axis by a small deviation angle is centered about the deviation angle with the same distribution as that of an undeflected electron about its trajectory angle. This situation is well known for TR, i.e the centroid of the far field radiation pattern "follows" the angle of trajectory of electron for forward TR and the specular reflection angle for backward (reflected) TR.

*Observations in the Detection Plane*

In addition to the angular coordinates $\theta_x, \theta_y$ it is convenient to introduce angular - cylindrical coordinates $\hat{\theta}, \varphi$ ($\theta_x = \hat{\theta}\cos\varphi$, $\theta_y = \hat{\theta}\sin\varphi$) to describe directions in the detection plane. In these coordinates the $\varphi$ plane of observation projected to the detection plane is represented by the line $\varphi = const$, the horizontal plane of observation by the line $\varphi = 0$ and the vertical plane of observation by the line $\varphi = \pi/2$. We will also use the vector $\vec{\theta}_e$ with components $(\theta_{xe}, \theta_{ye})$ and $\vec{\theta}_k$ with components $(\theta_x, \theta_y)$ to describe the direction of the trajectory of the particle and the direction of observation, respectively.

As it is shown above the center of the radiation pattern of any electron interacting with the mesh follows the direction of the trajectory of the particle. Mathematically, this means that the distribution of intensity produced by the particle with trajectory $\vec{\theta}_e$ can be written as:

$$I_{\|\perp}(\vec{r}_e,\vec{\theta}_k,\vec{\theta}_e) = I_{\|\perp}(\vec{r}_e,\vec{\theta}_k - \vec{\theta}_e) = I_{\|\perp}(\vec{r}_e,\theta,\varphi) \tag{13}$$

where $\theta, \varphi$ are the components of the vector $\vec{\theta}_e - \vec{\theta}_k$ in angular-cylindrical coordinates where $\theta = \sqrt{(\theta_y - \theta_{ye})^2 + (\theta_x - \theta_{xe})^2}$ and $\varphi$ is the angle between vector $\vec{\theta}_k - \vec{\theta}_e$ and the $\theta_x$ (horizontal) axis in the detection plane. The terms $I_{\|\perp}(\vec{r}_e,\theta,\varphi)$ are the patterns of radiation whose centroid directions are collinear to $\vec{V}$. In the code these terms are calculated for particles with trajectory angle $\theta_{xe} = 0$, $\theta_{ye} = 0$ and then used to calculate the pattern of radiation of particle with an arbitrary trajectory angle with respect to the $z$ axes. Functions $I_{\|\perp}(\vec{r}_e,\theta,\varphi)$ are calculated using formulas (8) and (10).

*Total radiation from two parallel foils*

In the interferometer the particle passes through two foils: (1) a perforated mesh and (2) a solid foil, producing DR and TR respectively. Using the variables $\vec{r}_e,\vec{\theta}_k,\vec{\theta}_e,\theta,\varphi$ the intensities parallel and perpendicular to the $\varphi$ plane of radiation can be written as combination of terms which depend on $\theta,\varphi$ and those which depend on $\vec{\theta}_k,\vec{\theta}_e$:

$$I_{\|}(\vec{r}_e,\vec{\theta}_k,\vec{\theta}_e) = I_{1\|}(\vec{r}_e,\theta,\varphi) + I_{2\|}(\theta,\varphi) + 2 I_{1\|}(\vec{r}_e,\theta,\varphi)^{1/2} \cdot I_{2\|}(\theta,\varphi)^{1/2} \cdot \cos \Psi(\vec{\theta}_k,\vec{\theta}_e)$$
$$I_{\perp}(\vec{r}_e,\theta,\varphi) = I_{1\perp}(\vec{r}_e,\theta,\varphi)$$
$$\tag{14}$$

where $I_{1\|}(\vec{r}_e,\theta,\varphi)$ and $I_{1\perp}(\vec{r}_e,\theta,\varphi)$ are the components of intensity of radiation from the first foil and $I_{2\|}(\theta,\varphi)$ is the component from the second foil and the total intensity is:

$$I_T(\vec{r}_e,\vec{\theta}_k,\vec{\theta}_e) = I_{\|}(\vec{r}_e,\vec{\theta}_k,\vec{\theta}_e) + I_{\perp}(\vec{r}_e,\theta,\varphi) . \tag{15}$$

Note that the interference phase $\Psi(\vec{\theta}_k,\vec{\theta}_e)$ does not depend on the coordinates of the particle and that the term $I_{1\perp}(\vec{r}_e,\theta,\varphi)$ does not participate in the interference but merely adds to the intensity "background". The exact expression for interference phase is given by:

$$\Psi(\theta_x, \theta_y, \theta_{xe}, \theta_{ye}) = \frac{kd \cos \nu}{\beta} \left[ 1 + \frac{\tan^2(\theta_{xe} - \nu)}{\cos^2 \theta_y} + \frac{\tan^2 \theta_{ey}}{\cos^2(\theta_x - \nu)} \right]^{\frac{1}{2}}$$

$$-kd \cos \nu \left[ 1 + \frac{\tan(\theta_x - \nu)\tan(\theta_{xe} - \nu)}{\cos^2 \theta_y} + \frac{\tan \theta_y \tan \theta_{ye}}{\cos^2(\theta_x - \nu)} \right] \quad (16)$$

$$\cdot \left[ 1 + \frac{\tan^2(\theta_x - \nu)}{\cos^2 \theta_y} + \frac{\tan^2 \theta_y}{\cos^2(\theta_x - \nu)} \right]^{-\frac{1}{2}} + \Delta\Psi$$

where $\Delta\Psi = \pi$ when the forward radiation from the first foil is reflected and interferes with the backward radiation from the second foil. In the limit $\theta_{xe,ye} \to 0$ and small $\theta_{x,y}$, the phase shown in Eq. 16, $\Psi = (kd/\beta)(1 - \beta \cos \theta) + \Delta\Psi \to d/L_V + \Delta\Psi$, and the two foil interference term (see Eq. (1)) reduces to the term $\sin^2(d/2L_V)$ given in Eq. 4.

The radiation produced by the scattered $S$ or unscattered $U$ fractions of the beam is a summation of radiations produced by all the particles from each beam fraction. The parallel component can be written as:

$$I_{\|S,U}(\vec{\theta}_k, \vec{\theta}_e) = T_{\|1}^{S,U}(\theta, \varphi) + T_{\|2}^{S,U}(\theta, \varphi) + 2\cos\Psi(\vec{\theta}_k, \vec{\theta}_e) \cdot T_{\|1,2}^{S,U}(\theta, \varphi) \quad (17)$$

where

$$T_{\|1}^{S,U}(\theta, \varphi) = N^{-1} \cdot \sum_{i}^{S,U} I_{\|1i}(\vec{r}_e^i, \theta, \varphi)$$

$$T_{\|2}^{S,U}(\theta, \varphi) = N^{-1} \cdot \sum_{i}^{S,U} I_{\|2i}(\theta, \varphi)$$

$$T_{\|1,2}^{S,U}(\theta, \varphi) = N^{-1} \cdot \sum_{i}^{S,U} (I_{\|1i}(\vec{r}_e^i, \theta, \varphi) \cdot I_{\|2i}(\theta, \varphi))^{1/2} \quad (18)$$

are summation terms collinear to the direction of the trajectory of the particle, with index $i$ representing a particular beam particle with coordinates $\vec{r}_e^i$ ($x_e^i, y_e^i$) within the beam cell. The summations are done for scattered $S$ particles (i.e. particles passing through the mesh wires) and unscattered particles $U$ (particles passing holes) separately.

The perpendicular component of the intensity is calculated in the same manner as described above:

$$I_{\perp S,U}(\theta,\varphi) = T_{\perp 1}^{S,U}(\theta,\varphi) = N^{-1} \cdot \sum_{i}^{S,U} I_{\perp 1i}(\vec{r}_e^i, \theta, \varphi) \qquad (19)$$

Note that the perpendicular components do not contain an interference phase term because the radiation intensity from the solid foil is TR and, as such, does not have a perpendicular component. The total radiation from the two interferometer foils produced by all particles of S or U fraction with trajectory $\vec{\theta}_e$ is then:

$$I_{TS,U}(\vec{\theta}_k, \vec{\theta}_e) = I_{\parallel S,U}(\vec{\theta}_k, \vec{\theta}_e) + I_{\perp S,U}(\theta, \varphi) \qquad (20)$$

In practice these summations shown in Eqs. (18,19) are only done for 24 values $\varphi_m = m \cdot 15^0$, $m = 0,1,2,...23$ and a few tens of points $\theta_l$ in the interval $0 \leq \theta_l \leq 6/\gamma$ for the scattered and unscattered beam fractions. This data is saved in a Table and used later to determine additionally needed values by interpolation.

*Computing the effect of beam divergence*

The effect of beam divergence on the intensities computed above is performed by means of a two dimensional angular convolution. In order to perform this convolution it is necessarily to know the intensity produced by all particles of each beam fraction (scattered and unscattered) with trajectory angle $\vec{\theta}_e$ at the observation point $\vec{\theta}_k$, as well as the distribution of trajectory angles of the beam electrons.

We model the distribution of electron trajectory angles as a sum of up to three individual Gaussian components. For instance in the case of the mesh the wires substantially scatter electrons up to few mrads completely "hiding" the original angular distribution which is usually a fraction of one mrad. The scattered portion of the beam evidently has wider angular distribution than the unperturbed beam passing through the holes and is represented as a single wide Gaussian component. However, the angular distribution of the "unperturbed" unscattered portion of the beam is more complex and cannot be represented by a single Gaussian distribution. We have modeled this situation by splitting the U fraction of the beam into two individual Gaussian components, the minimum number required for our fits. Note that the zero angle of the total distribution is the same before and after scattering and is the same for all components.

In angular coordinates and using the small angle approximation, the multi-Gaussian component beam can be presented as:

$$\frac{dN(\theta_{xe}, \theta_{ye})}{d\theta_{xe} d\theta_{ye}} = \sum P_n(\theta_{xe}, \theta_{ye}) = \sum A_n \exp\left(-\frac{\theta_{xe}^2}{2\sigma_{xn}^2} - \frac{\theta_{ye}^2}{2\sigma_{yn}^2}\right) \quad (21)$$

where $n$ is the number of Gaussian components including scattered and unscattered portions, $\sigma_{xn}$, $\sigma_{yn}$ are standard angular deviations and $A_n$ are normalization constants. In this paper, the numbers $n = 1$ and 2 designate the 1st and 2nd components of the unscattered beam and $n = 3$ designates the single scattered component.

The radiation produced by the $n^{th}$ component at the observation point $\theta_x, \theta_y$ is obtained by integrating over the phase space area $-q\sigma_x \leq \theta_{xe} \leq q\sigma_x$, $-q\sigma_y \leq \theta_{ye} \leq q\sigma_y$ where $q = 3$ is usually a sufficiently large limit for the integration:

$$J_n(\theta_x, \theta_y) = \int P_n(\theta_{xe}, \theta_{ye}) I_n(\theta_x, \theta_y, \theta_{xe}, \theta_{ye}) d\theta_{xe} d\theta_{ye} \quad (22)$$

where

$$I_1(\theta_x, \theta_y, \theta_{xe}, \theta_{ye}) = \tau_1 I_{TU}(\vec{\theta}_k, \vec{\theta}_e) \quad (23)$$

$$I_2(\theta_x, \theta_y, \theta_{xe}, \theta_{ye}) = \tau_2 I_{TU}(\vec{\theta}_k, \vec{\theta}_e) \quad (24)$$

$$I_3(\theta_x, \theta_y, \theta_{xe}, \theta_{ye}) = I_{TS}(\vec{\theta}_k, \vec{\theta}_e) \quad (25)$$

and $\tau_1$, $\tau_2$ ($\tau_1 + \tau_2 = 1$) are the relative weights of the Gaussian fractions of the unscattered beam components.

As described above the code first calculates the two dimensional $(\theta, \varphi)$ distributions of the radiation components $T_{\|1}^{S,U}(\theta_l, \varphi_m)$, $T_{\|2}^{S,U}(\theta_l, \varphi_m)$, $T_{\perp1}^{S,U}(\theta_l, \varphi_m)$ and cross terms $T_{\|1,2}^{S,U}(\theta_l, \varphi_m)$. According to the convolution procedure the intensity in the direction $\theta_x, \theta_y$ is a sum of intensities weighted by the distribution of electron trajectories angles.

Figure 4. shows the beam angular distribution as a shaded area in the detection plane $\theta_x, \theta_y$. The dark line in the Figure represents the observation plane for a particular group of electrons (scattered or unscattered) in the distribution, in which the perpendicular and parallel intensities are calculated. From these intensity components we can calculate the contribution of a particular group of particles to the total intensity. By repeating this procedure for all the groups of electrons in the distribution the total intensity can be calculated and compared to measured total intensity.

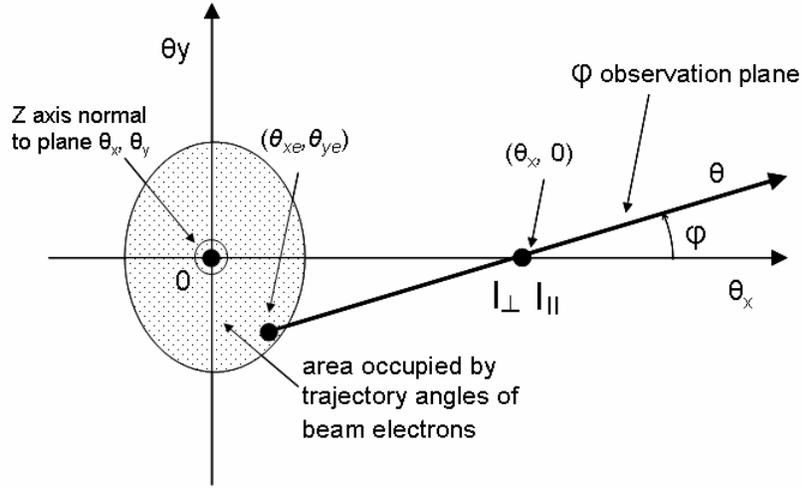

Figure 4. Schematic of the observation plane, showing the angular space occupied by the beam electrons and the $\theta, \phi$ scan direction for a single electron trajectory angle in this distribution.

The intensity produced by a group of electrons with trajectory $\theta_{xe}, \theta_{ye}$ in the direction $\theta_x, \theta_y$ is calculated by the following way: 1) the values $\theta, \varphi$ are calculated, where $\theta = \sqrt{(\theta_y - \theta_{ye})^2 + (\theta_x - \theta_{xe})^2}$ and $\varphi$ is the angle between vector $\vec{\theta}_k - \vec{\theta}_e$ and the $\theta_x$ (horizontal) axis; 2) the phase term $\cos\Psi(\vec{\theta}_k, \vec{\theta}_e)$ is calculated using Eq. (16); 3) the terms given in Eqs. (23, 24, 25) are calculated using saved data [see Eqs. (18,19) and the discussion following Eq. (20) above] and linear interpolation in the rectangle $\theta_l \leq \theta \leq \theta_{l+1}$, $\varphi_m \leq \varphi \leq \varphi_{m+1}$ as required; 4) the terms $P_n(\theta_{xe}, \theta_{ye}) I_n(\theta_x, \theta_y, \theta_{xe}, \theta_{ye})$ are calculated for each of the n components and the integrations of these functions using Eq. 22 are performed. Finally the code calculates the horizontal and vertical scans of intensity produced by all fractions of the beam. The calculated scans are used to compare and fit the experimental scanned data to the calculated scans.

*OTRI Limit*

In the case of an OTR interferometer which consists of two *solid* foils, the parallel component of TR produced by the beam with trajectory angular components $\theta_{xe}, \theta_{ye}$ in the observation direction $\theta_x, \theta_y$ is given analytically by Eq. (5) where $\theta = \sqrt{(\theta_y - \theta_{ye})^2 + (\theta_x - \theta_{xe})^2}$ and the detection plane is coplanar to $\vec{k}, \vec{V}$. In this case Eq. (22) reduces to:

$$I_n(\theta_x, \theta_y) = 2\int P_n(\theta_{xe}, \theta_{ye}) I_{TR}(\theta)(1 + \cos\Psi) d\theta_{xe} d\theta_{ye} \tag{26}$$

which is the OTRI limit. In our analysis of OTRI data, the angular convolution for TR interferometer is performed using Eq. (26) and two Gaussian components to represent the unscattered beam distribution.

*Additional Convolutions*

To account for variations of the beam energy and the observation wavelength, we can optionally perform separate convolutions or averages over these variables as well as angular convolution. To account for finite band of observed wavelengths, we assume that the spectral characteristic of the band pass filter is rectangular and perform an convolution of the intensity scan over the filter transmission function. To account for possible variations in beam energy we can perform a convolution of the scanned intensity over energy under the assumption that the energy distribution has a cosine distribution. Optional forms for the energy variation are Gaussian or rectangular distributions.

*Unpolarized OTRI and ODTRI*

In the experiments described below unpolarized OTRI and ODTRI images are used, i.e. no polarizer was used. The reason for this is that analysis shows that the use of a polarizer does not give any advantage when a 2D angular convolution code is used to evaluate the data.

On the other hand, calculations show that the *polarized intensity* is less sensitive to the corresponding perpendicular angular divergence $\sigma_\perp$. For example, if the polarization axis is the *x* direction, $\sigma_\perp = \sigma_y$. If $\sigma_\perp \ll \sigma_x$, the divergence component along the polarization axis, there will be little difference between polarized and non polarized intensities. However, at moderate and large values of $\sigma_x$, $\sigma_\perp$ should be taken into account in order to correctly calculate the polarized intensity. This means that a 2D angular convolution must be done in any case. Hence there is no advantage in using the polarizer. For these reasons we measure and use the total intensity interferogram obtained without the use of a polarizer. From this interferogram the divergence in any direction can be determined by simply scanning along the desired angular direction.

**Demonstration of Simulation Code Results**

In Figure 5. we present results of simulation code calculations of the sum of ODR contributions from the beam electrons which pass through the holes of the mesh (unscattered beam) and the sum of ODR contributions from the electrons passing through the mesh wires (scattered beam), for a 5 micron thick, 750 lines per inch copper mesh, with 25 μm square holes and a cell period *p* = 33 μm (55% transparency). The percentage of the ODR intensities from unscattered and scattered beam components is about 10% of the total radiation. The OTR generated at the mirror from the scattered and unscattered components are 55% and 45% respectively, in accordance with the mesh transparency. The beam energy used in the calculations shown in Figure 5. is 50 MeV. Similar calculations were done for 95 MeV. The code results show that the angular distributions of ODR from electrons passing through the holes and the wires are similar to that of OTR from a solid foil. Since all these distributions are slowly varying function of observation angle, the main effect of beam divergence, represented mathematically by a convolution of the intensity (see Eqs. 1 and 2) with a distribution of electron angles is to blur or reduce the visibility of the interference fringes. This effect is the basis of beam divergence diagnostics with both OTRI and ODTRI when the energy spread is smaller than the normalized divergence which is the case in the present study.

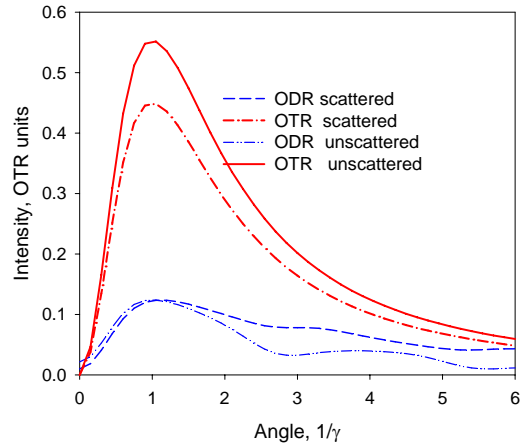

Figure 5. Computer simulation of the parallel components of intensities of the ODR and OTR from the scattered and unscattered components of the copper micromesh; tilt angle of the plane of observation φ = 0; particle trajectory is parallel to the *z* axes.

The interference term for OTRI and ODTRI is the same since this term depends only on the relative phase of the radiation from the first foil and the mirror. Figure 6. shows the interferences generated from the individual ODR intensity components described above. Note that for the scattered component the fringe visibility is close to zero, i.e. the fringes produced by the scattered component of ODR is completely washed out. This is intentionally done by choosing the atomic number and thickness of the mesh for a given beam energy, such that heavy scattering ($\sigma_s = 4$ mrad) of the electrons ensues.

In this situation the fringes due to the unscattered (unperturbed) beam component are made visible above the smooth (incoherent) scattered beam contribution. The two components add to form the black curve in Figure 6. The fringe visibility is affected by the inherent (unperturbed) beam divergence, which for illustration is $\sigma = 0.5$ mrad. The wavelength chosen for the calculation is 650 nm with a delta function band pass.

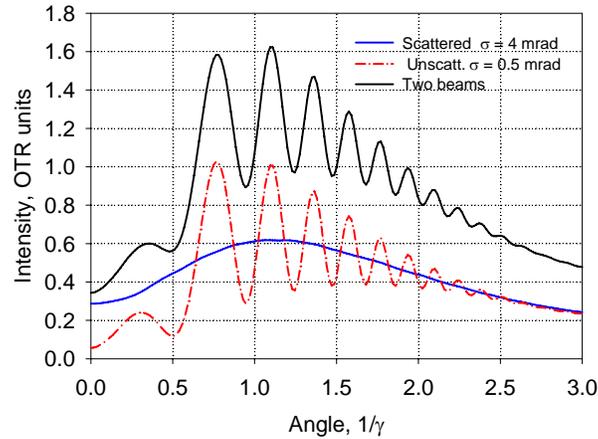

Figure 6. Interferences produced by unscattered (red ) and scattered (blue) ODR components from the mesh with OTR from the mirror and their sum (black).

**Description of the Experimental Setup**

The beam energies used in our experiments are 50 and 95 MeV. The setups for both experiments are essentially the same. Both employ optical trains which accept and maintain an angular field of view of approximately $10/\gamma$ and transport the light to cameras positioned away from the foil-mirror position to reduce the x-ray background. A schematic of the optics used at the BNL/ATF is given in Figure 7. for illustration. Details of the experimental setup of the NPS 95 MeV experiment have been previously described in [19]. For each experiment, care must be taken to insure that the second camera is focused on the plane of the mirror, i.e. the OTR radiator/reflector. This is done with the help of a graticule target, whose surface is coplanar with the mirror.

To image the far field angular distribution, i.e. the OTR or ODTR interference pattern we used Apogee Instruments Inc., 16 bit, Peltier cooled, high QE, low noise CCD cameras each of which is equipped with an electronically controlled mechanical shutter;

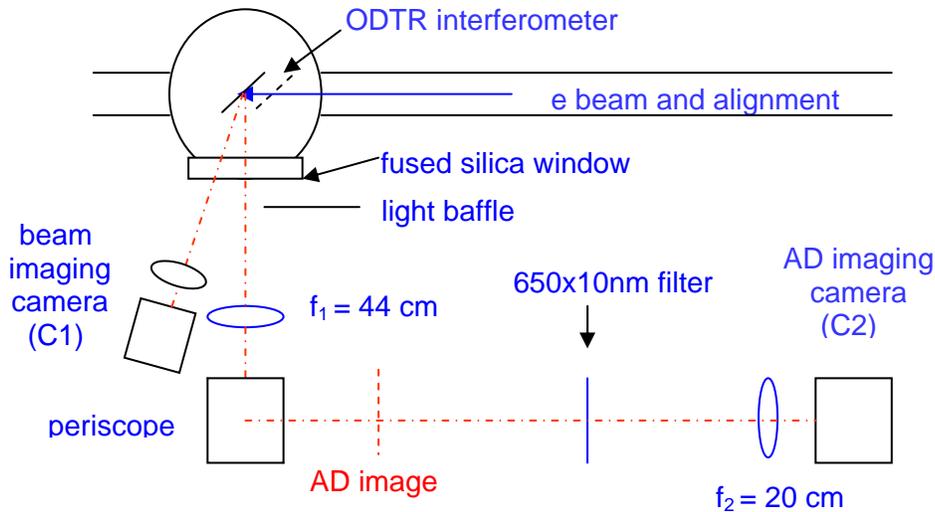

Figure 7. Top view of experimental setup at BNL/ATF showing beam line, vacuum vessel, interferometer and optics.

this allows the CCD to integrate the light produced from multiple electron beam pulses. A model Alta E47+ was used at ATF and a model AP230E was used at NPS. A second less sensitive RS 170, 8-bit CCD camera (Cohu 4912 or GBC-CCTV 500E) was used to monitor the beam's spatial distribution.

The first lens shown in Figure 7, which has a focal length f1=44cm, is placed 47 cm from the mirror. In the focal plane of this lens an image of the far field angular distribution (AD) appears. A second lens whose focal length $f_2$ = 20cm placed at 168 cm from the mirror is used to re-image the AD in the image plane of camera C2. The object and images distances for C2 are 77 and 28 cm respectively. A light baffle is used to prevent direct reflection from the first foil or mesh from entering the optical path.

Identical interferometers but with different foil-mirror spacings: d = 37 and 47 mm, for the 50 and 95 MeV beams respectively, were used in the experiments.

A photograph of the target ladder housing the interferometers is shown in Figure 8. This apparatus was mounted on a stepper motor driven, 6-inch linear actuator. One of four positions (components) of the ladder could be placed into the beam: 1) a graticule, used to determine the magnification of the system; 2) an aluminized Silicon mirror cut from a 0.5 mm wafer, used alone for alignment of the optics with an upstream laser; 3) the OTR interferometer, consisting the mirror and a 0.7 micron thick foil of 99.5% pure aluminum mounted on a stretcher ring (the apparatus seen on the right hand side of

Figure 8. and also in reflection from the mirror); and 4) the ODR interferometer consisting of the mirror and a micromesh foil, which is also mounted on a circular stretcher ring.

The foils and mirror are parallel and tilted at 45 degrees with respect to the direction of the electron beam. The forward directed radiation from each foil and the backward OTR are observed in reflection from the mirror through a fused silica view port. To align and focus the far field camera, a HeNe laser (632nm) pointing down stream along the beam line axis was used to create an optical diffraction pattern with the micromesh in place. The resulting diffraction pattern formed a cross of dots, with the central dot (zero$^{th}$ order) specifying the direct beam. The higher order dots were located at angular positions $\theta = n\lambda/p$ where $n$ is the diffraction order, $\lambda$ is the wavelength and $p$ is the hole period. This pattern provided an excellent angular calibration source for the far field camera.

The ATF linac at Brookhaven National Laboratory and the Naval Postgraduate School linac beam have the following characteristics: ATF: 1.5 pps, 500-700 pC per pulse, NPS: 60 pps, 0.1-0.8 µA average current; the normalized rms emittances have been previously measured to be $\varepsilon^n_{rms} \sim 1$ and $\sim 200$ mm mrad respectively; the measured energy spreads are $\Delta\gamma/\gamma < 0.5\%$ for ATF and $\Delta\gamma/\gamma < 5\%$ for NPS; and the focused beam sizes used in our experiments were approximately 100 microns and 1000 microns with corresponding normalized rms beam divergences $\gamma\theta_{rms} = 0.05$ and $0.10$. The foil-mirror spacings were determined from calculations to produce an optimal number of interferences for the beam energy and expected range of divergence for each beam. The optimal spacings and band passes for these two situations were determined prior to the experiment by applying the results of computer code runs for both OTRI and ODTRI interferences.

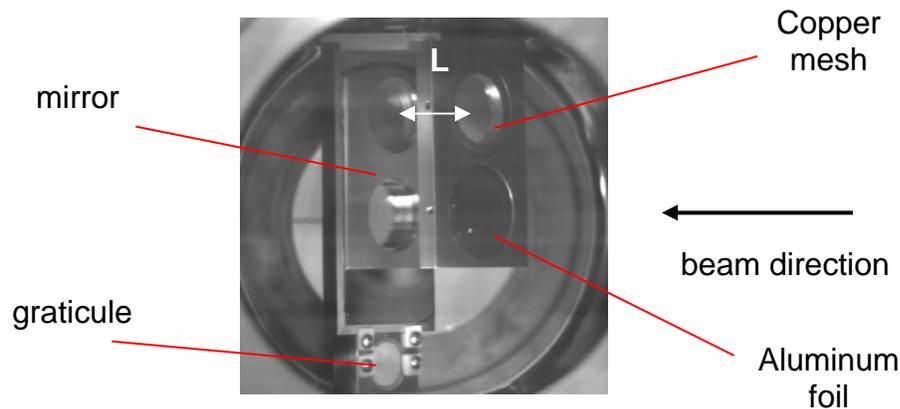

Figure 8. Target Ladder showing ODTR and OTR Interferometers, mirror and graticule.

At NPS a pair of quadrupole magnets up stream of the target chamber were used to magnetically focus the beam to either an *x* or *y* waist condition at the mirror position. The waist condition was confirmed by observing the maximum sharpness of the higher order interference fringes [20]. At a beam waist the visibility of the observed OTR or ODTR interference fringes in the *x* (horizontal) or *y* (vertical) directions is a measure of the corresponding x or y rms beam divergence. Thus, together with knowledge of the rms (x or y) size at the corresponding (*x* or *y*) waist obtained from the spatial image of the beam and the corresponding rms divergence obtained from the interference pattern, the rms *x* and *y* beam emittance could be determined.

**Results and Analysis**

*Data Fitting Procedure*

ODTRI and OTRI experiments were performed on the NPS 95 MeV accelerator focusing the beam to both *x* and *y* waist conditions at the site of the interferometer mirror and on the ATF accelerator for two different beam tunes. A camera focused on the mirror monitored the beam size in both cases. To obtain a good signal to noise ratio (S/N>2) we found it necessary to integrate over many beam pulses to build up a good interferogram. At the NPS this time was about 60 seconds, while for the ATF the integration time was of the order of 5 minutes.

For each interferogram we extracted two mutually perpendicular scans, e.g. horizontal and vertical. Averaging of the intensity over an angular sector about the horizontal or vertical direction is first performed at each radial distance from the center of the pattern. The sector angle is chosen such that the visibility of the fringes along a *sector averaged* line scan through the center of the pattern is not noticeably different by the eye from that of a simple un-averaged, albeit noisy, single line scan through the center. Sector averaging improves the signal to noise ratio substantially especially for noisy images and provides smooth line scans which are then fit to the simulation code calculations to give the value of the rms divergence

To fit the scanned data to simulation code scans we compare the data to a family of theoretical curves each obtained for a particular set of beam parameters: divergence, energy, energy spread and fractional weight, when more than one beam component is required, and interferometer foil spacing. The goal is to achieve the best set of parameters which simultaneously provides a 'best fit' to both the horizontal and vertical sector averaged data scans.

The goodness of the fit of measured $E(\theta)$ and calculated $T(\theta)$ scans is determined first qualitatively by eye and then more rigorously using the following procedure. The essence of this procedure is to scale the data scan by a constant $A$ until the best fit of the data scan to the theoretically calculated scan is obtained. To do this we have written a code to compare the similarity of the two functions $A \cdot E(\theta) \geq 0$ and $T(\theta) \geq 0$ in the interval $(\theta_1, \theta_2)$ defined in terms of the integral RMS deviation defined as:

$$D(A) = \frac{1}{(\theta_2 - \theta_1)} \left[ \int_{\theta_1}^{\theta_2} \left( \frac{A \cdot E(\theta) - T(\theta)}{A \cdot E(\theta) + T(\theta)} \right)^2 d\theta \right]^{1/2} \tag{27}$$

where $A \cdot E(\theta)$ is the experimentally measured intensity distribution, scaled by $A$, an arbitrary amplitude, and $T(\theta)$ is the line scan calculated from the simulation code described above. The closer the functions $A \cdot E(\theta)$ and $T(\theta)$ are to each other the smaller the value of $D$; the further these functions are from each other the closer $D$ is to unity.

We define the maximum similarity of the two functions when D(A) is at its minimum value, i.e. $D(A_{min}) = \min(D(A))$. Any change of the set of the parameters used to calculate $T(\theta)$ changes the shape of theoretical curve and the values of $D_{min}$ and $A_{min}$. The goal is to achieve the best set of parameters which simultaneously gives the best similarity for both the horizontal and vertical scans. The "best fit" occurs when we have found a set of beam parameters, interferometer parameters and values of $D_{min}$ and $A_{min}$ for which the eye judges that the best similarity is achieved. Note that in the ideal case $D_{min} = 0$.

Adjustment of the parameters used to fit the experimental scan data to simulation code or theoretical calculations is performed in the following way:

1) parameters of the interferometer and expected (starting) parameters of the beam, i.e. foil separation $d$, filter pass band, electron energy band, parallel $\sigma_p$ and normal $\sigma_n$ angular divergences, angular interval for fitting are inputted.
2) theoretical/simulation code calculations are performed for both horizontal and vertical scans and the theoretical and renormalized experimental curves are plotted.
3) a comparison of experimental and theoretical curves and adjustment of the input parameters is made to achieve the best similarity between theory and experiment simultaneously for both horizontal and vertical scans, i.e. by minimizing the rms deviation function D.
4) adjustment of $\sigma_p$, $\sigma_n$ energy and $d$ are made to get the best fit to the interference pattern.
5) a check of the effect of energy spread and pass filter on fringe visibility is done; if these effects are negligible, fine tuning of parameters is then done to minimize both the rms deviations for horizontal and vertical scans.
6) if needed, the beam distribution is split into two fractions and adjustment of the parameters of second beam is performed to improve the best fit.
7) if necessary, a third beam fraction is introduced and adjustment of the parameters of the third beam is performed. (NB: this third component is only used to estimate effect of the *scattered component* from the mesh wires).

*Example of Data Fitting Procedure*

To illustrate the procedure employed to fit the sector averaged line scans used in all our analyses of OTRI and ODTRI patterns, we will use an OTRI from NPS as an example. The OTR interference pattern for the NPS beam focused to a *y* (vertical ) beam waist is shown in Figure 9. The measured rms y size of the beam at the *y* waist condition is ~ 1mm. The OTRI pattern was obtained by exposing the far field CCD camera for 45 seconds; the picture is taken with an optical filter band pass filter in place ( 650nm, 70 nm FWHM band pass). The colored sectors overlaying the image in Figure 9. show the angular regions used to average the intensity at each radius measured from the center of the pattern.

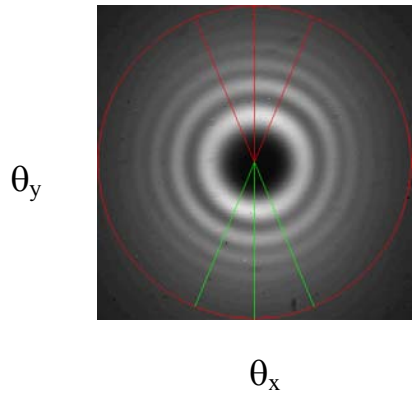

Figure 9.   OTR interferences for the 95 MeV NPS at a *y* (vertical) waist; overlay: sectors over which the intensity at each radius is averaged.

Figure 10. shows the fit to the vertical line scan of OTRI taken at a *y* using the convolution of a 2D Gaussian function with Eq. (1), for two different values of the rms beam divergence, $\sigma = 0.6$ mrad and 0.7 mrad, along with experimental data, i.e. the sector averaged vertical scan of Figure 9.  The overall best fit to the data (i.e. all the fringes) is seeming provided by the $\sigma = 0.7$ mrad fit. However, note that the best fit to the higher order fringes (i.e. angles larger than $1.5/\gamma$ shown in the expanded region on right of Figure 10.) is better with a value of $\sigma = 0.6$ mrad. On the other hand this value produces a fit that is poorer for the lower order fringes , i.e. angles smaller than $1.5/\gamma$.

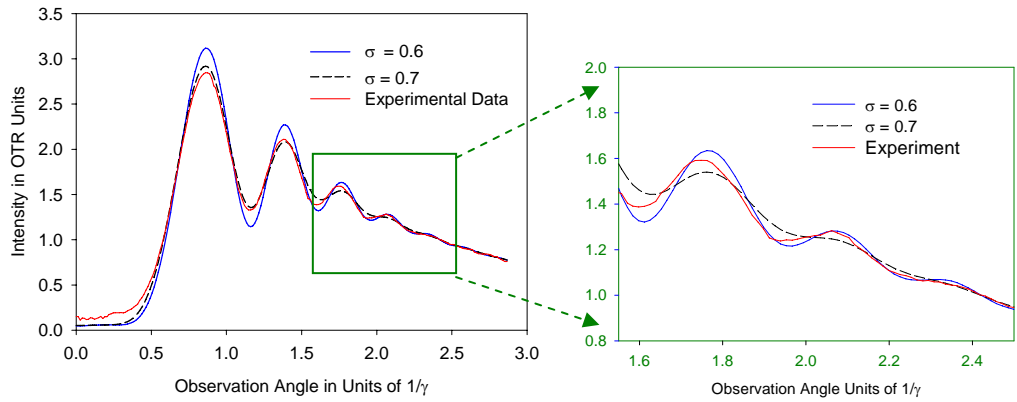

Figure 10. Comparison of the effect of single Gaussian distribution functions with different rms widths on the OTRI fringes (left); expanded plot region (right).

A variational analysis the interference phase term in Eq. (1) shows that the effect of divergence on the fringe visibility is proportional to the fringe order so that the effect of increasing the divergence is seen as a reduced fringe visibility for the higher order fringes first [21]. The higher order fringes are better fit by single Gaussian with $\sigma = 0.6$ mrad but the lower order fringes are not fit well with this same function; this is evidence that the real beam angle distribution is not well represented by a single Gaussian.

To improve the fit to the data we have introduced a second 2D Gaussian function in addition to the primary Gaussian to model the distribution of electron angles. The fractional amplitudes and rms widths of both Gaussians were adjusted to provide the best fit (dot dashed blue line) to the data (solid red line) shown in Figure 11. The primary distribution fraction is weighted by 0.75 and its rms width, $\sigma = 0.6$ mrad.

The effect of the primary distribution (Comp1) on the OTRI fringes is shown by the dashed black curve. The effect of the secondary distribution (Comp2) is shown by the dot-dashed green line. The total effect of the two components is represented by the dot-dashed blue line. As is seen from Figure 11. the overall fit to the data is excellent with the two component model over the entire range of observation angles.

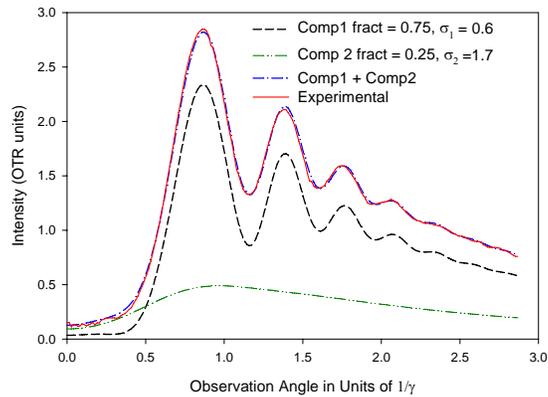

Figure 11.  Comparison of effect of convolution of two weighted Gaussians
with different fractional amplitudes and rms widths on OTRI fringes.

Similarly we used a two component distribution to represent the angular distribution of the unscattered electrons to fit the ODTRI data. However, in the case of ODTRI a third Gaussian component representing the scattering of the beam in the wires of the mesh, which is always present regardless of the number of inherent beam components, is also used in the fit. This component is similar in its effect to the broad primary beam component shown in Figure 11.

*NPS Data Fits*

An ODTRI pattern at the *y* waist NPS beam condition obtained with an integration time of 60 seconds and the same band pass filter used for the OTRI is shown in Figure 12. (left) along with a vertical line scans of the pattern on the left and the multi-component Gaussian best fit to the data (right).

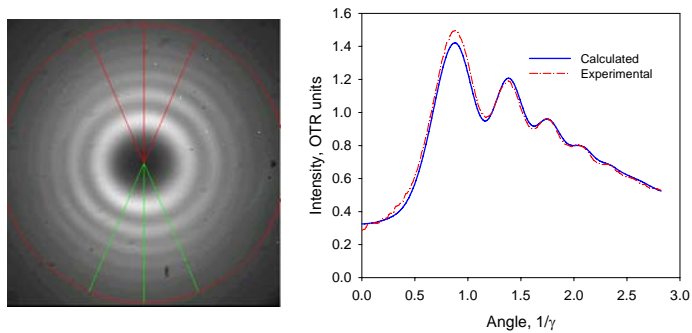

Figure 12.  ODTRI pattern (left) and sector average vertical line scans (right).

The best fitted values for the *y* component of the beam divergence from the OTRI and ODTRI averaged line scans are 0.58 mrad and 0.56 mrad respectively, showing a consistent value for the divergence of the primary beam component from the two independent measurements.

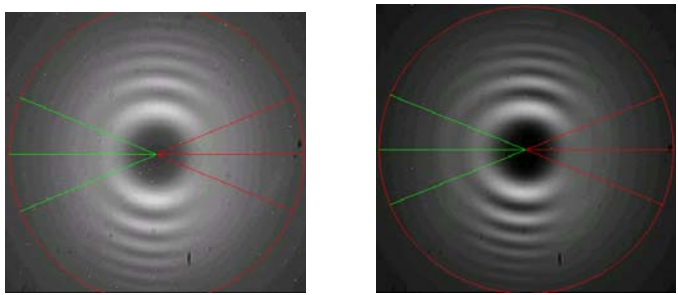

Figure 13.  ODTR (left) and OTR (right) interference patterns at an x waist condition; overlay:sectors of the angular regions over which the intensity is averaged to produce an x line scan.

Figure 13. presents ODTRI and OTRI for the NPS beam focused to an *x* (horizontal) waist condition.  These pictures show a lower visibility of the fringes in the horizontal or *x* direction in comparison to the higher visibility of *vertical* fringes as seen in Figures 9. and 12., indicating that the x (horizontal) beam divergence is larger than the *y* (vertical) beam divergence.

Figure 14. presents fits to the horizontal sector average line scans obtained from the interference patterns presented in Figure 13. The *x* (horizontal) divergence obtained from fitting both the OTRI and ODTRI averaged line scans is 1.2 mad. This value is about twice as large as the *y* (vertical) divergence given above.  The quality of the *x* waist

fits is obviously not as good as the *y* waist fits because of the lower signal to noise present in the θ_x direction of the interferogram.

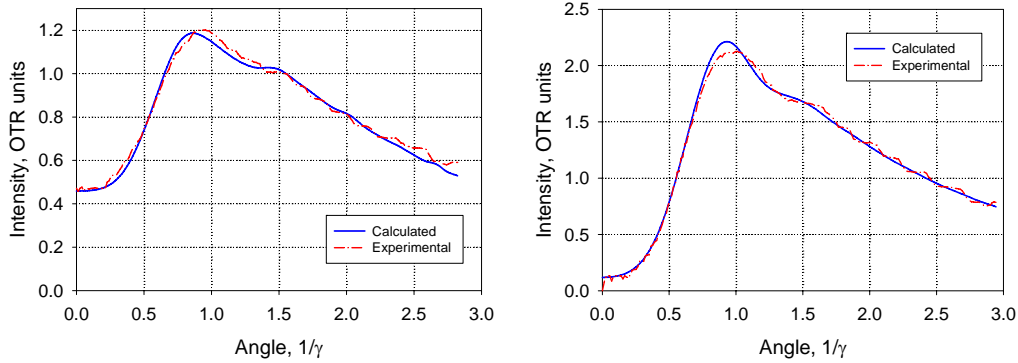

Figure 14. Horizontal averaged line scans of ODTRI (left) and OTRI (right) shown in Figure 13.

A comparison of the other ODTRI and OTRI fit parameters is provided in Table 1. There is a slight difference in the spacing, 36.5 mm for the ODTRI vs. 37.2 mm for the OTRI, which is most likely due to a small difference in orientation of the two foils with respect to the beam direction due to rotational wobble in the linear drive. Previous analysis [15,21] has shown that the position of the fringes is a sensitive function of the beam energy and spacing but that the visibility of the fringes is primarily affected by the divergence, when the energy spread of the beam is small in comparison to the normalized divergence.

Table 1. Fitted beam parameters for NPS beam Y and X waists.

| Waist | Method | Scan | Energy (MeV) ±0.2 | Comp1 (%Tot) ±5% | $\sigma_1$ (mrad) ±5% | Comp2 (%Tot) ±5% | $\sigma_2$ (mrad) ±10% | d (mm) ±0.2 |
|---|---|---|---|---|---|---|---|---|
| Y | OTRI | Vert. | 93.5 | 72 | 0.58 | 28 | 1.4 | 37.2 |
| Y | ODTRI | Vert. | 93.5 | 69 | 0.56 | 31 | 1.5 | 36.5 |
| X | OTRI | Horiz. | 93.5 | 100 | 1.2 | | | 37.2 |
| X | ODTRI | Horiz. | 93.5 | 100 | 1.2 | | | 36.5 |

This is the case for NPS since $\Delta\gamma/\gamma \approx 0.03$ and $\gamma\sigma \approx 0.12$. Our code calculations verify this and show that even if $\Delta\gamma/\gamma = 0.1$ for NPS there would have be little effect on the fringe visibility.

*ATF Data Fits*

ODTRI-OTRI experiments were done at the ATF accelerator for two different beam tunes, i.e. two sets of beam sizes and divergences, which were obtained by tuning a magnetic triplet upstream of the ODTRI interferometer. The beam parameters for each beam tune are independently determined from multiple screen beam size measurements and transport code calculations. Three beam profile monitors (YAG screens) were placed upstream of the ODTR interferometer and one beam profile monitor (fluorescence screen) downstream. The electron beam size at each monitor was measured. By fitting the beam sizes with a trajectory calculated by transfer matrices of quadrupoles and drift spaces, the sigma matrix at the interferometer position was computed and correspondingly the beam size, divergence and emittance were obtained as well. The parameters for the first beam tune were: x = 0.18 mm, y = 0.27 mm, $\sigma_x$ = 0.31 mrad and $\sigma_y$ = 0.22 mrad.

Figure 15. shows ODTR and OTR interference patterns obtained with the first beam tune. The ODTRI and OTRI patterns are obtained with integration times of 480 and 360 seconds respectively with a 650 x 10 nm band pass filter. The narrower band pass, i.e. 10 nm for ATF vs. 70 nm for NPS, is required to obtain the sensitivity (greater number of fringes) required to measure the lower divergence of the ATF beam. The smaller band pass and the additional lower average current of the ATF in comparison to NPS necessitated a longer integration time for the ATF, which was limited by the build up of background due to x-rays. Consequently the use of sector averaging was especially important for the ATF data.

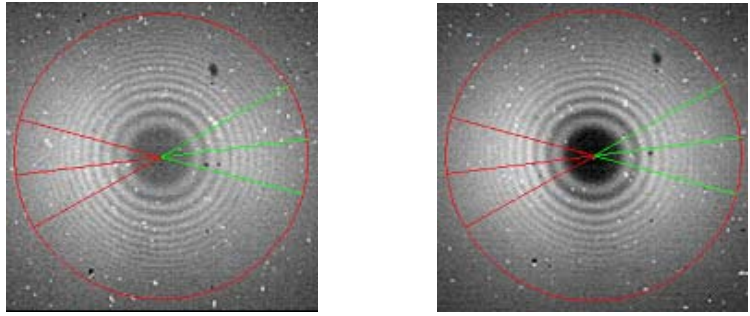

Figure 15. ODTR (left) and an OTR (right) interference patterns for ATF Tune 1 with overlay of horizontal sectors used to average the fringe intensity.

Note the apparent offset of the colored sectors from horizontal, which is due to a slight rotation of the mirror with respect to the optical axis. This offset is observable and known from the diffraction pattern of the laser, which follows the same optical path as the ODTR.

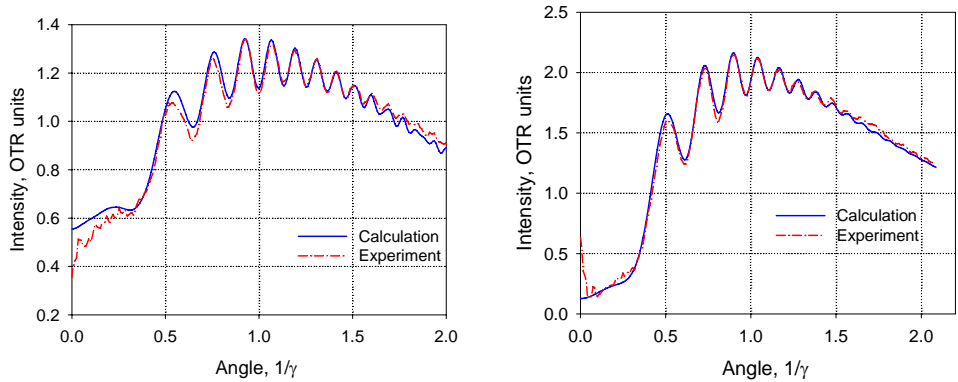

Figure 16. Sector averaged line scans of ODTRI (left) and OTRI (right) from Fig.15.

Horizontal sector averaged line scans along with theoretical fits are shown in Figure 16. Note that the number of visible fringes in the ODTRI scan exceeds the number in the OTRI scan. This is expected since the visibility of ODTRI is not affected by scattering in the first foil.

Figure 17. shows the ODTR and OTR interference patterns obtained for the second beam tune and Figure 18. shows the corresponding sector average line scans. For this tune the beam parameters are: x = 0.18 mm, y = 0.15 mm, $\sigma_x$ = 0.37 mad and $\sigma_y$ = 0.75 mrad.

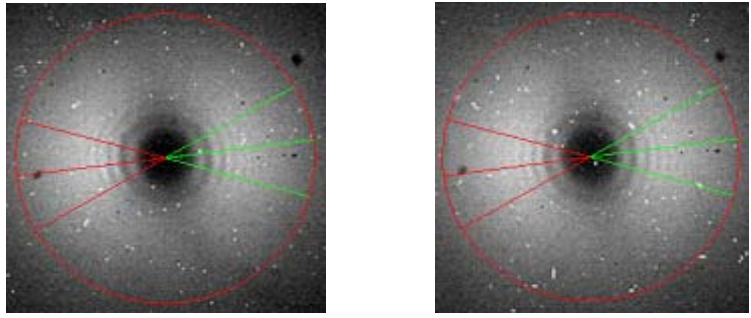

Figure 17. ODTRI (left and OTRI (right) for the second beam tune of the ATF linac.

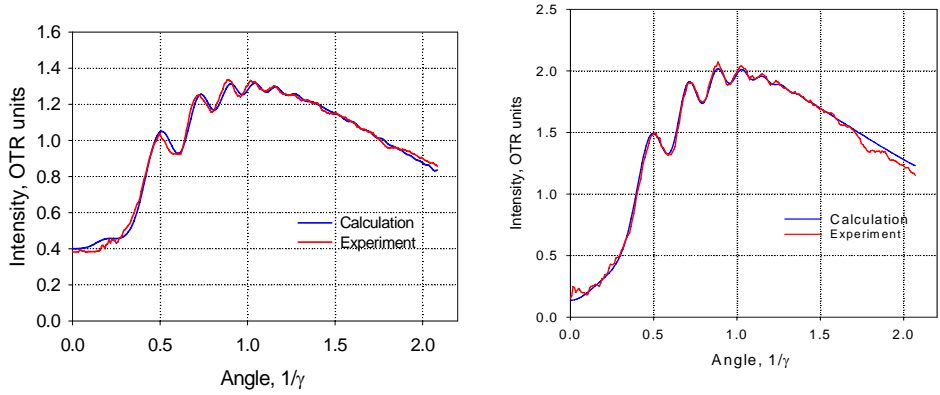

Figure 18. Averaged line scans of ODTRI (left) and OTRI (right)corresponding to Fig.17.

Table 2. Fitted beam parameters for ATF beam tunes 1 and 2.

| Tune | Method | Scan | Energy MeV ±0.2 | Comp1 % Tot ±5% | $\sigma_1$ mrad ±5% | Comp2 % Tot ±5% | $\sigma_2$ mrad ±5% | d mm ±10% | $\sigma_E$ mrad ±0.5 |
|---|---|---|---|---|---|---|---|---|---|
| 1 | OTRI | H | 50.7 | 28 | 0.35 | 72 | 1 | 47 | 0.31 |
| 1 | OTRI | V | 50.7 | 38 | 0.3 | 62 | 1 | 47 | 0.22 |
| 1 | ODTRI | H | 50.0 | 33 | 0.28 | 67 | 1 | 44.5 | 0.31 |
| 1 | ODTRI | V | 50.0 | 55 | 0.28 | 45 | 1 | 44.5 | 0.22 |
| 2 | OTRI | H | 50.3 | 335 | 0.5 | 67 | 1.6 | 47 | 0.37 |
| 2 | OTRI | V | 50.3 | 33 | 0.75 | 67 | 1.6 | 47 | 0.75 |
| 2 | ODTRI | H | 49.3 | 33 | 0.4 | 67 | 1.6 | 44.5 | 0.37 |
| 2 | ODTRI | V | 49.3 | 33 | 0.65 | 67 | 0.8 | 44.5 | 0.75 |

A complete set of fitted parameters for the two beam tunes at ATF is given in Table 2. The narrow Gaussian distribution full width, i.e. the rms divergence of the primary beam $\sigma_1$, should be compared to the divergence $\sigma_E$ obtained with the multiple screen - transport code measurements. Again, the smallest normalized divergence, i.e. $\gamma\sigma \approx 0.3$ is still much less than the measured energy spread for ATF, i.e. $\Delta\gamma/\gamma \approx 0.005$. This is also verified by code calculations which show that an energy spread of up to 2% has little effect on the fringe visibility.

**Discussion**

We have examined the possible causes of the inferred bimodal distributions and consequent two beam divergences obtained from our fits to the ODTRI and OTRI data. These are listed and analyzed below.

1. Energy Spread

The energy spread of the ATF beam was monitored during the experiment and is less than 0.5%. Both variational analysis of the interference terms in Eq. 1. [21] and our convolution codes show that this spread is too small to be responsible for the observed fringe blurring; i.e. the energy spread would have to be 16 times larger (8%) to show the effect observed at ATF. The energy spread at NPS is higher than ATF, i.e. a few percent. However, the divergence of the NPS beam is also higher. Both variational analysis and computer convolution calculations show that the energy spread is not sufficient to significantly affect the observed visibility.

2. Bandwidth of the Filter

A fixed bandwidth optical bandpass filter was used in all runs, so blurring due to changes in wavelength outside the bandpass is not possible. Numerical convolution of the transmission functions for the filter used at both ATF ( 650 x 10 nm FWHM ) and NPS (650 x 70 nm FWHM) shows in each case that the filter bandpass has a negligible effect on the fringe visibility.

3. Beam Halo

We have considered the possibility that there is a beam halo component in addition to a core component and that the beam core and halo have different spatial and angular distributions. The presence of a halo is certainly possible at NPS and in fact dark current component is known to exist and has been observed in previous experiments, although its divergence has not been measured. Dark current components have been observed in other linacs also, e.g. the 8 MeV ANL-AWA. The present analysis shows that this component is at the 20% level for NPS. Such a component would not likely be noticeable, e.g. from an observation of the beam spatial distribution, which is limited by the dynamic range of the beam imaging camera (8 bit CCD) used.

However, at ATF, the presence of a large (i.e. 60% of total) background beam component, inferred from both the OTRI and ODTRI fits, would probably have been previously observed but this has not been reported. Since our observations of the beam profile was again limited to 8 bits, we cannot completely rule out the existence of a halo in our runs. However, since a large halo component is unlikely, we have examined another possible explanation for the inferred bimodal distribution and large second component fraction, i.e. the effect of beam stability during the rather long integration times required for ATF experiments (360s for OTRI and 480s for ODTRI).

*4.* Beam Instabilities

There are several types of beam stabilities that could be present and possibly affect our results. These include:

a. Jitter in the beam position

This type of jitter has no effect on the far field *angular distributions* of OTRI and ODTRI.

*b.* Random jitter in the trajectory angle of the beam

This effect would combined with the effect of the beam core divergence resulting in as a single Gaussian distribution whose full width would be calculable from quadrature addition of the FWHMs of the components, one related to the inherent temperature of the beam, the other to a jitter of the trajectory angle would be seen as a single Gaussian with a wider rms width: $\sigma_{tot}^2 = \sigma_b^2 + \sigma_{jitter}^2$. A single Gaussian with width σ$_{tot}$ would have a predictable effect on the fringe visibility. Since the fringes cannot be fit with a single Gaussian distribution, a random jitter effect must be ruled out.

    c. Non-random jitter of the trajectory angle of the beam

This effect would appear as a distinct non Gaussian distribution and possibly explain the need for a second distinct component in addition to the core beam angular distribution. Since we did not continuously monitor the beam position during the experiment and acquire the data that would allow a statistical calculation of the jitter, we cannot rule out this possibility. We therefore conclude that a nonrandom instability in the beam position, during the long image integration times, is a possible explanation for the observed two Gaussian component fit. A future experiment will be needed to test this possibility.

**Conclusions**

We have obtained nearly identical divergences and beam fractions using two independent measurements, i.e. OTRI and ODTRI. The divergence obtained agree wel with those obtained by independent multiple screen measurement-transport code calculations. The analysis of the OTRI and ODTRI data are very different. OTRI analysis uses a direct convolution of an exact analytic expression to obtain a fringe pattern, which is then fit to the data. The analysis and fitting of ODTRI, on the other hand, is much more complicated and requires a simulation code to perform. It is very unlikely that these two independent calculations and fits to data would produce nearly identical results that both agree with the independent multiple screen analysis. We conclude that ODTRI and OTRI are indeed being affected by the same, real physical effect and that the simulation code and data fitting procedures we have employed are correct and consistent.

We emphasize that we have not set out to prove that electron beams, under certain operating conditions can show bimodal distributions - which is the conclusion of our analysis of both OTRI and ODTRI. Rather, we have set out to show that ODTRI is a viable diagnostic technique to measure beam divergence *whatever* the beam conditions.

Our results show that the divergences and component intensities measured by ODTRI match those obtained by OTRI, for two separate experiments on two different accelerators with significantly different beam properties. Furthermore, the divergences due to the core beam component are in agreement with other independent measurements and simulation code results. Thus, we have well demonstrated that ODTRI is a valid new divergence diagnostic method which extends and can even replace OTRI as a diagnostic for low energy and or very low divergence beams.

Since the ODTRI fringes are sensitive to the *unperturbed* beam, which passes through the holes of a micromesh foil, ODTRI overcomes the lower limit on measurable divergence present in a conventional OTR interferometer, i.e. scattering in the solid first foil. Hence ODTRI can be used to diagnose lower emittance and/or lower energy beams.

Our demonstration of ODTRI as a useful rms divergence diagnostic indicates that it may be possible to use ODTRI to make *localized* beam divergence and trajectory angle measurements, i.e. within the beam's spatial distribution, and therefore to produce a transverse phase space map of the beam. Optical phase space mapping (OPSM) has already been demonstrated using OTRI [19,20]. Presuming the problem of the lower intensity yield of ODR compared to OTR can be overcome, e.g. by increasing the integration time or beam current, OPSM using ODTRI should be straightforward.

**Acknowledgement**

This work is supported by the Office of Naval Research and the DOD Joint Technology Office.